\documentclass{osa-article}

\journal{oe}

\articletype{}

\usepackage{lineno}

\begin{document}

\title{Multi-resonant open-access microcavity arrays for light matter interaction}

\author{Thomas H. Doherty,\authormark{1} Axel Kuhn\authormark{1,2} and Ezra Kassa\authormark{1,3*}}

\address{\authormark{1}Department of Physics, University of Oxford, Parks Road, OX1 3PU Oxford, United Kingdom \\
{\authormark{2}axel.kuhn@physics.ox.ac.uk} \\
\authormark{3}Experimental Quantum Information Physics Unit, Okinawa Institute of Science and Technology Graduate University, 1919-1 Tancha, Onna, Kunigami, Okinawa, 904-0495, Japan}
\email{\authormark{*}ezra.kassa@oist.jp} 


\begin{abstract*}
We report the realisation of a high-finesse open-access cavity array, tailored towards the creation of multiple coherent light-matter interfaces within a compact environment. We describe the key technical developments put in place to fabricate such a system, comprising the creation of tapered pyramidal substrates and an in-house laser machining setup. Cavities made from these mirrors are characterised, by laser spectroscopy, to possess similar optical properties to state-of-the-art fibre-tip cavities, but offer a compelling route towards improved performance, even when used to support only a single mode. The implementation of a 2$\times$2 cavity array and the independent frequency tuning between three neighbouring sites are demonstrated. 
\end{abstract*}


\section{Introduction}
Fabry-P\'{e}rot optical cavities are a platform for many landmark scientific advances in atomic, molecular and optical physics, from the resolution of atomic spectral lines to the development of laser resonators. In recent years, single emitters have been strongly coupled to the vacuum mode of an optical cavity, granting the ability to perform coherent quantum control at the single particle level~\cite{Boca:04}. This allows one to implement a range of quantum techniques, with applications in quantum computation, communication and enhanced sensing~\cite{Kimble:08}. Key examples include on-demand single-photon generation~\cite{Kuhn:02,Keller:04}, atom-photon gates~\cite{Reiserer:14} and photon-photon gates~\cite{Hacker:16}. These schemes can even be expanded to universal quantum computing, based on a network of many emitter-cavity nodes~\cite{Ritter:12}. Each of these demonstrations of controlled cavity quantum electrodynamics has been predicated on significant advances in optical cavity design and manufacture.

{}{Arrays of high-finesse cavities have already been demonstrated as a means of scaling quantum systems \cite{Dolan:10, Derntl:14, Wachter:19, Jin:22}. However, to the authors' knowledge, multi-resonant cavity arrays have yet to be demonstrated.}
{}{Whilst arrays of high finesse cavities have been produced by lithographic techniques \cite{Dolan:10, Derntl:14, Wachter:19, Jin:22}, here, we employ CO$_2$ laser machining as has first been used} in the creation of fibre-tip Fabry-P\'{e}rot cavities (FFPCs)~\cite{Hunger:10}, to produce multi-faceted mirror templates on bulk substrates that eventually allow for bringing multiple cavities into simultaneous resonance. {}{Another leading approach in optical cavity engineering is the use of nanophotonics based resonators as pointed out in}\cite{Chang:2018}. {}{Such resonators have already been coupled to neutral atoms} \cite{Thompson:2013} {}{but are intrinsically incompatible with ion traps \cite{Harlander2010, Brownnutt2015}}.

{}{In contrast,} FFPCs have been integrated with a broad range of the most promising qubit candidates for scalable quantum information processing, including ions~\cite{Takahashi:20}, neutral atoms~\cite{Gallego:18}, nitrogen vacancies~\cite{Albrecht:13} and quantum dots~\cite{Sanchez:13}. Fibre-tip mirrors are shaped by laser ablation to provide highly curved surfaces and, upon application of a highly reflective dielectric coating, demonstrate optical losses in the parts per million regime \cite{Hunger:10}. These mirrors can be used to construct cavities of a small mode volume, for strong coupling between emitter and photons, open-access to the cavity mode for emitter localisation and an intrinsic overlap to a fibre guided mode for photon extraction. However, despite these traits leading to the widespread adoption of FFPCs~\cite{Pfeifer:22}, a range of unresolved issues regarding scalability remain. 

We consider limits to cavity scalability in two broad classes: those which impact the generation of distributed entanglement across a quantum network and those that add complexity to the fabrication or operation within a laboratory environment. With regard to the first concern, a key metric of performance for a cavity-based network interface is its photon extraction efficiency, which, depending on the specific network architecture, will place a limit on the rate of heralded entanglement between spatially separated emitters~\cite{Gao:21}. Photons generated within an FFPC are inherently coupled to a fibre-guided mode, providing direct connectivity to an optical network. However, in using single-mode fibres to preserve spatial coherence, photon extraction efficiencies are commonly limited to the order of 5\%, due to the mismatch between the cavity mode and fibre-guided spatial mode~\cite{Gallego:16}. This efficiency has been greatly improved by the introduction of gradient index lenses to the fibre-tip ~\cite{Gulati:17},
{}{achieving mode-matching efficiencies up to 89\% for a 400$\mu$m long FFPC. This is comparable to the efficiencies of coupling the output of substrate cavities to single mode fibres~\cite{Ritter:12}}

To effectively operate a network of optical cavities, the constraints and resources required to continuously run individual nodes must be minimal. A key consideration is mechanical stability, as for a high finesse resonator to maintain a fixed resonance frequency, its length must be stable better than $\lambda/(2\mathcal{F})$ where  $\mathcal{F}$ is the finesse at wavelength $\lambda$. This stability requirement typically evaluates to the picometer scale. Accordingly, the cavity design should ensure that any mechanical resonances are decoupled from environmental noise to allow the cavity length to be persistently stabilised by electronic feedback~\cite{Gallego:16}. This also requires careful consideration of any mechanical contacts between the cavity and surrounding vacuum chamber, such as feedthroughs for optical fibres~\cite{Peerzada:20}. Despite these susceptibilities, highly stable systems have been demonstrated various times over the past decades, for example, using miniature cavities made from bulk mirrors \cite{Stefan:2005}, or using fibre-tip mirrors in glass ferrules ~\cite{Saavedra:21}, which form a quasi-monolithic structure.

{}{The tapered substrates used in this work are akin to other bulk-mirror substrates used in cavity Quantum Electrodynamics experiments, which showed favourable properties for both photon extraction and mechanical stability~\cite{Ritter:12, Nguyen:17}. Since photon capture is performed using free-space optical elements, a mechanical bridge is avoided between vacuum chamber and cavity, improving the decoupling of mechanical resonances from environmental noise. The use of macroscopic optics for cavity-mode matching offers the ability to achieve photon capture into a fibre with efficiencies approaching that of spliced GRIN lenses in an FFPC \cite{Gulati:17}. However, macroscopic optics offer a degree of flexibility and potential reconfigurability over fiberized approaches, at the cost of compactness and additional alignment procedures. Polarisation is generally well preserved with macroscopic optics, with scope to assist the fidelity in polarisation-encoded photonic entanglement schemes.} However, unlike super-polished mirror cavities, the use of laser ablation creates strongly curved facets which gives rise to a highly confined cavity mode. The resulting small mode volume enhances the atom-cavity cooperativity, without moving towards the concentric regime by increasing cavity length. Indeed, tapered mirror cavities offer a compelling option in high finesse cavity construction, even when supporting only a single cavity mode.

{}{Here, we describe the manufacture, implementation and characterisation of compact arrays of high-finesse optical cavities, as a proposed route towards scalable light-matter interfaces in quantum networks. In using laser ablation to machine an array of mirror features on the end facet of tapered substrates, FFPCs are the natural benchmark and comparison to this work. However, the production of cavity mirror arrays, suitable for mediating interactions with single quantum emitters, has been well demonstrated using a variety of glass shaping techniques. A notable example is focused ion beam milling, able to create highly curved and tightly spaced mirror features on bulk substrates \cite{Dolan:10}. Due to the presence of molten glass layer inherent to laser ablation in this regime, it is likely unable to achieve arrays of similar compactness. However, by virtue of the surface tension effects present in this layer, these mirrors are more readily able to achieve the high optical quality required for low cavity loss.}

\begin{figure}[h!]
\centering\includegraphics[width=13cm]{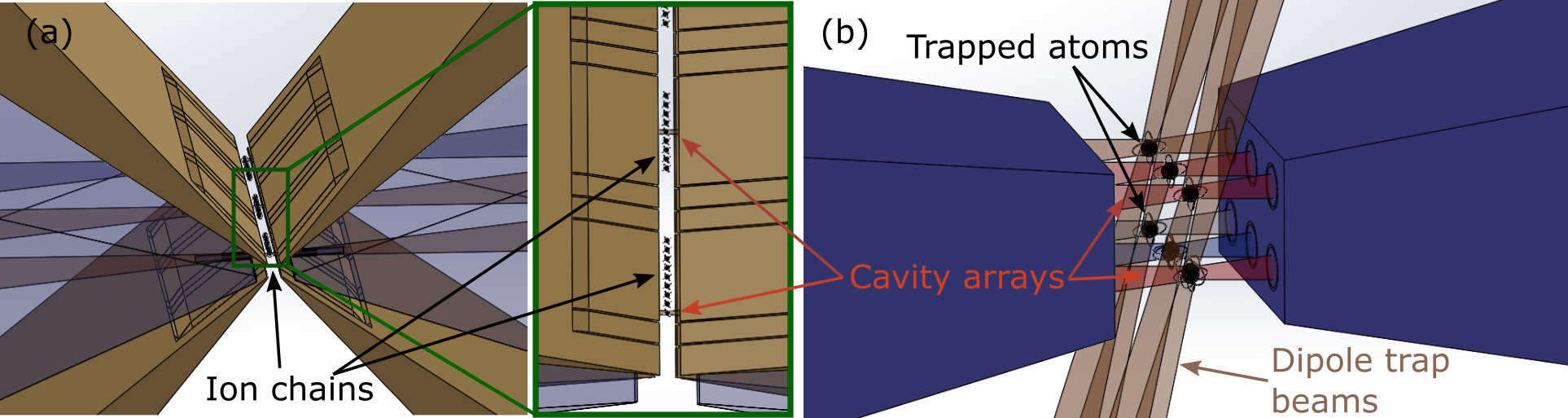}
\caption{
Artist's view illustrating the possible combination of cavity arrays with either (a) chains of trapped ions in a segmented ion trap or (b) a grid of atoms held in optical tweezers. Tapered mirror substrates are shown in blue and co-resonant cavity modes are highlighted in red. Trapping electrodes in (a) are shown in gold, and dipole trapping light in brown in (b). Up to three cavities can be brought into simultaneous resonance in a triangular arrangement. Thus two co-resonant ion-cavity systems might be realised in combination with linear ion chains, whereas three co-resonant atom-cavity systems are feasible using more versatile trapping geometries. We note that other cavity modes (shown in different colours) could be used for stabilising the overall length of the array.  
} 
\label{fig_prospects}
\end{figure}

We expect that compact cavity arrays on a bulk substrate can be combined with existing multi-emitter trapping methods to realise scaled network node architectures, of which two prospective examples are shown in Fig.~\ref{fig_prospects}. The first is a chain of ions within a segmented linear Paul trap~\cite{Nigmatullin:16}, with each trapping zone coupled to an individual cavity mode. Given the compact geometry of the tapered substrate, distortion of the trapping potential by the dielectric surfaces will be suppressed~\cite{Podoliak:16}. Alternatively, a grid of neutral atoms may be localised within a square array of cavity modes, using an array of optical tweezers~\cite{Stuart:18}. In both systems, multiple atom-cavity interfaces have been illustrated within the same vacuum chamber, whose size and complexity is on scale with a single interface. Length stabilisation of the entire array can be performed using a single reference cavity because all of the cavities reside on the same pair of substrates. 
{}{These envisaged experimental schematics further motivates our use of conventional Fabry--P\'{e}rot cavities for this work. Where, in comparison to approaches in nanophotonics which also report small mode-volume and open access, a simple pair of facing mirrors provides flexibility with respect to intra-cavity trapping method.}

In the following, we describe our experimental platform and process for laser ablation, tailored towards the fabrication of micro-mirror arrays on a range of different substrates. We detail the production of single and multi-feature tapered mirrors and determine the quality of our cavities by laser spectroscopy and consider routes towards future improvement. Finally, in Section \ref{sec:MultiFeatures}  we demonstrate an operational array of four cavities.


\section{Implementation and characterisation of mirror arrays}\label{Sec:TapMirFab}

\subsection{Laser Ablation System}
Laser ablation is a widely employed process in the fabrication of open-access optical cavity mirrors, used to carve near-spherical depressions into a glass substrate~\cite{Hunger:12}. It requires intense laser irradiation to induce localised evaporation, leaving a concave feature suitable for the creation of a cavity mirror of tight curvature. Low scattering losses are ensured by sustaining laser illumination for a period of milliseconds, establishing a molten layer which is smoothed by the action of surface tension. The interplay of the evaporative and molten processes is largely governed by the characteristics of this laser exposure sequence, allowing one to shape the mirror geometry. However, as a thermal process it is notoriously difficult to both predict and finely control, thereby requiring ablation systems to support the rapid characterisation of fabricated surface topographies and allow the empirical derivation of desired pulse sequences. 

\begin{figure}[h!]
\centering\includegraphics[width=12cm]{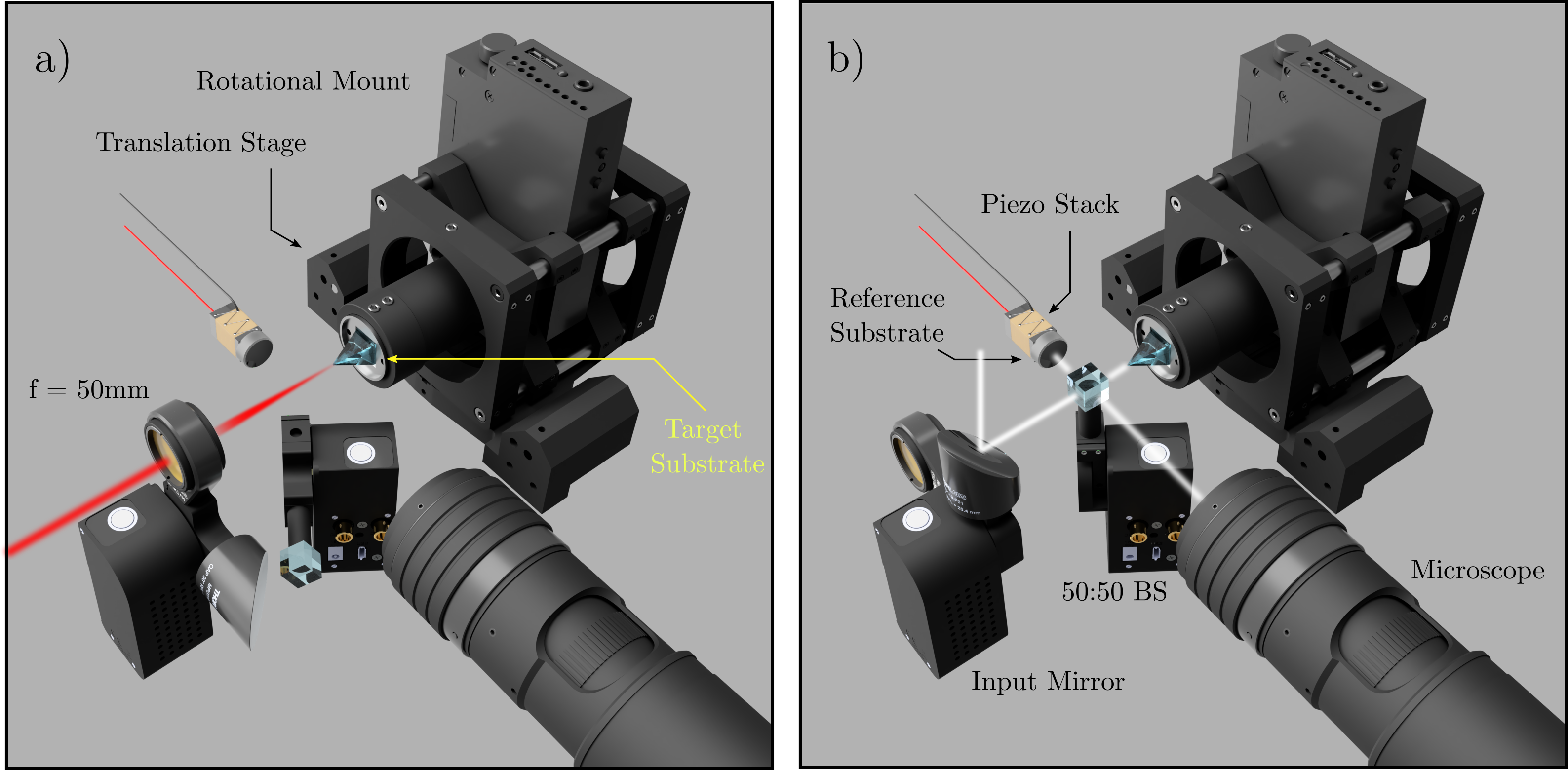}
\caption{Schematic of the laser ablation and interferometric surface characterisation setup. a) For laser ablation, a 10.6~$\mu$m laser beam is focused to a spot size of $\approx$50$~\mu$m on the target substrate, which is mounted on a rotational and translational stage. Pulse trains of peak optical power of 1-4~W and duration of 5-150~ms are applied to the substrate, to achieve the desired mirror geometry. b) A Michelson interferometer is used for surface reconstruction with white light illumination. This uses a 50:50 beam splitter and a planar reference substrate. The optical path lengths to the reference and target substrates are tuned by piezoelectric stacks, with their joint interference pattern measured by a microscope.}
\label{fig_ablate}
\end{figure}

The laser ablation system developed for the fabrication of our cavity mirrors is based upon an architecture reported by Takahashi et al., used in the fabrication of fibre-tip mirrors with markedly low ellipticity~\cite{Takahashi:14}. In broad similarity with the reported system, our apparatus uses a 10.6~$\mu$m CO$_2$ pulsed laser focused to a beam radius of approximately 100~$\mu$m onto the target substrate. Circular polarisation is ensured to reduce feature ellipticity, albeit with an inherent ellipticity of 0.6 remaining in the beam profile itself. Pulse shaping is performed by modulating the duty cycle of the laser, creating exposure durations of 5-150~ms of optical powers of 1-4~W. Fluctuations in the laser power are reduced to $<0.5$\% by the inclusion of an optical isolator and with the implementation of electronic feedback to the laser. However, persistent variation was observed in fabricated feature geometries, which was notably reduced by constructing each depression from many repeated exposures. 

\begin{figure}[h!]
\centering\includegraphics[width=6cm]{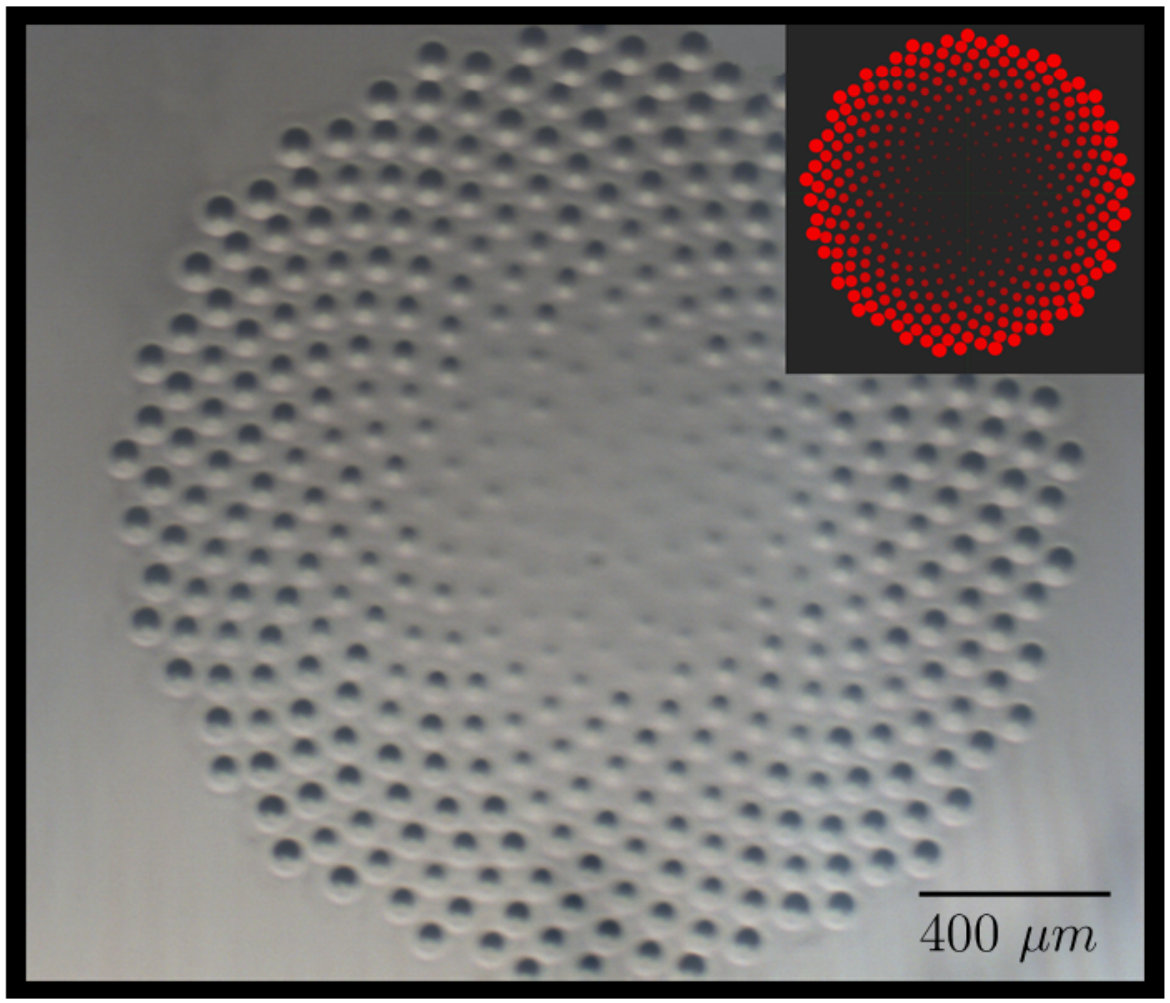}
\caption{An optical microscope image demonstrates that precise mirror patterns can be constructed by autonomous control of the stepper motors, here showing approximately 400 features in a phyllotactic arrangement. The inset shows a drawing of the intended pattern.}
\label{fig_phylotax}
\end{figure}

To provide support for tapered mirror substrates, key alterations were made to the Takahashi approach, particularly for substrate mounting and surface reconstruction. Schematics of this beamline are shown in Fig.~\ref{fig_ablate}. This apparatus accepts a broad range of geometries for both glass substrates and optical fibres, mounted by bespoke adaptors into a $\varnothing$1" cylindrical fixture. In turn, the target substrate is installed into a stepper motorised rotation mount (Thorlabs K10CR1/M), used to set the facet orientation of the target substrate with respect to axis of ellipticity in the addressing laser. This is secured onto a six-axis translation stage (Thorlabs MAX609L/M \& DRV001), including stepper motors and closed-loop piezoelectric transducers in each Cartesian axis. Stepper motor translations are applied in the plane of the laser propagation axis to build arrays of features, similar to~\cite{Petrak:11}, over a region of 4~mm~$\times$~4~mm. An example of such an array, fabricated on a standard planar $\varnothing$1/2" cylindrical substrate, is shown in Fig.~\ref{fig_phylotax}. 

Employing the procedure reported in~\cite{Ott:16}, each ablation feature is constructed from a pattern of laser exposures, with piezoelectric translation performed between each shot. By careful shaping of the micro-pattern, this method is demonstrated to enhance feature diameter, tailor ellipticity and bring fabricated features closer to a spherical profile. In our system the maximum pattern size is 30~$\mu$m~$\times$~30~$\mu$m, limited by the range of the piezoelectric translation stage. {}{By way of example, for the $93.2\mu$m diameter mirror discussed in Section~\ref{sec:mirrors}, the micro-pattern consisted of 81 ablation sites repeated 5 times, with exposure power of 3.8W and duration increasing from 120ms to 130ms.} 

For rapid surface reconstruction, the infrared beamline could be electronically switched to a white light Michelson interferometer by motorised flip mounts (Thorlabs MFF101/M). This allowed the target substrate to remain stationary between ablation and reconstruction, with a transition time of $<1$~s. Following standard techniques~\cite{Degroot:15}, the surface profile of the target substrate is measured in the static interference pattern produced with a planar reference substrate. For enhanced resolution, the profile can also be obtained by scanning the path length of one interferometric arm and observing the coherence envelope imposed by the illumination source. This gives a measure of feature curvature, diameter and ellipticity, each crucial in the production of cavity geometries of choice. 


\subsection{Tapered Substrate Design and Processing}

{}{Bulk substrates} allow the cavity mode to freely expand into the {}{glass} so that it can emerge from the substrate with its polarisation and transverse mode profile preserved to then be manipulated with external mode-matching optics. This facilitates the injection of laser light for cavity stabilisation {}{and the} capture of intra-cavity generated photons. {}{To allow for better compatibility of such mirrors with atom or ion traps, the substrates are often coned or tapered such that the end facet is 0.1-0.3 mm in diameter, which is comparable to the end facet of an optical fibre.}  The taper angle must be sufficiently shallow to contain the emanating cavity mode, while permitting open-access for emitter loading and persistent localisation. For neutral atoms, this may require support of a tightly focused optical dipole trap, without excessive illumination of the cavity mirrors. In this case, the taper angle must be sharp and the dimensions of the facet restricted. For ions, perturbation of the trapping potential by the dielectric mirror surfaces creates a limit on trapping lifetime, thus imposing a similar motivation to constrain the size of the substrate facet~\cite{Podoliak:16}. For the case of a cavity array, a further consideration of the facet size comes in the layout and quantity of the desired features.

\begin{figure}[h!]
\centering\includegraphics[width=12cm]{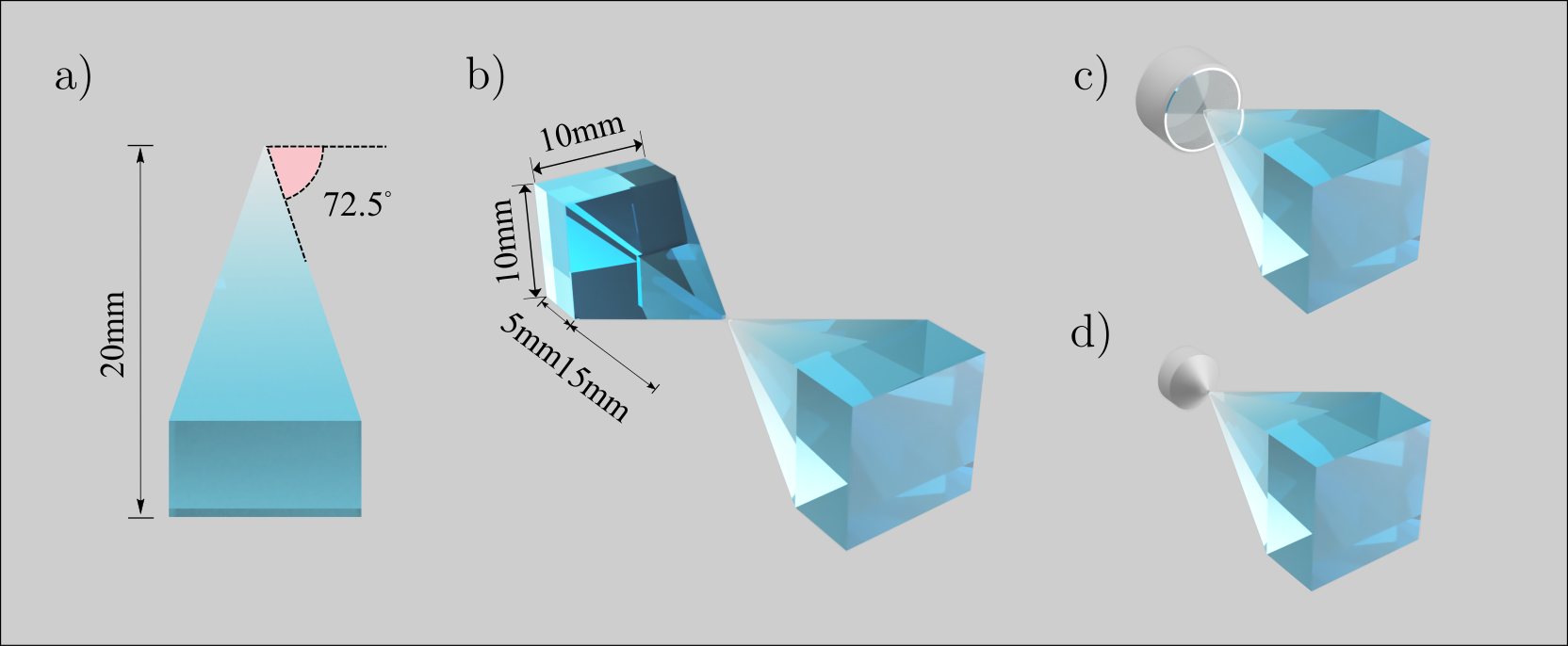}
\caption{a) A basic schematic of a pyramidal substrate is shown, highlighting its exterior dimensions and taper angle. The pyramidal substrates are shaped from UV-Fused silica, with a weight of 2.2~g and glass volume of 1000~mm$^3$. b) One could form a cavity using a pair of pyramidal substrates, as illustrated. This allows for a facet size on the order of 100~$\mu$m or larger, similar to an optical fibre-tip. {}{c) The pyramidal mirrors were characterised using a near-planar reference mirror of known optical properties. d) A super-polished substrate can be machined to improve optical access when paired with a tapered ablated substrate.}
}
\label{fig_pyramid}
\end{figure}

\begin{figure}[h!]
\centering\includegraphics[width=12cm]{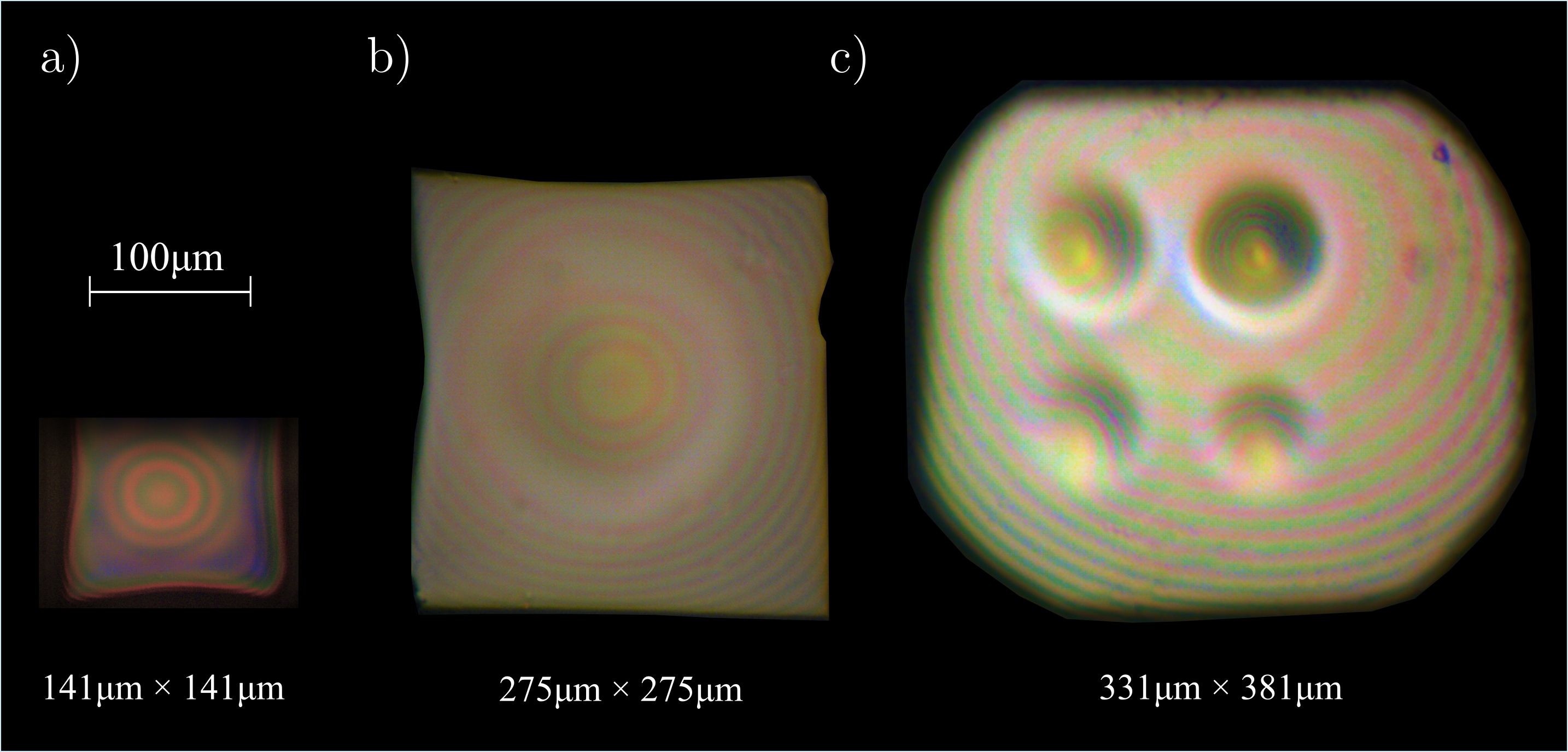}
\caption{Phase-interferograms of three pyramidal substrate tips post-laser machining, representing different facet sizes and mirror arrangements. a) The smallest facet produced was 141 µm $\times$141 µm, however, displayed some warping towards the feature edge. This was attributed to surface tension and  heat transport during the ablation process, where the spherical feature is modified by the proximity of rectangular facet edges. Accordingly, it would be challenging to produce larger features on this facet, or further decrease facet size. The feature shown has mean radius of curvature 790~$\mu$m, 1/e Gaussian diameter of 62.7~$\mu$m and ellipticity of  $\varepsilon=0.46$. b) When the substrate was expanded to 275~$\mu$m $\times$ 275~$\mu$m, significantly less warping was visible towards the facet edge. The shown feature has mean radius of curvature 1191.9~$\mu$m, diameter 93.2$\mu$m and ellipticity $\varepsilon=0.56$. c) An example array of features is shown, on a large facet diameter. At this scale, edge effects could be largely ignored, with care taken to provide sufficient separation between neighbouring features. Several spherical geometries were produced on this feature by modification of the laser pulse power.}
\label{fig_facets}
\end{figure}

For this work, a range of cavity mode geometries and array layouts were explored, requiring a conservative taper angle and the ability to tailor the facet size. Fig.~\ref{fig_pyramid} shows the tapered substrates, with interferometric images of sample facets given as Fig.~\ref{fig_facets}. The substrates were shaped from UV-Fused Silica by a commercial optics manufacturer (Knight Optical UK Ltd.), leaving a pointed tip on the substrate. We polished this tip to the desired facet geometry using a motorised random orbital plate and 20 nm grit paper. Eventually the surface was subjected to thermal annealing inherent to laser ablation in order to ensure a satisfactory surface roughness of approximately 3\r{A}. The experimental verification of this is given in Section~\ref{sec:mirrors}.

Performing ablation on the tapered mirror substrates introduced experimental complexities not observed with fibre-tip facets. The geometry of the generated ablation features was observed to depend both on the chosen exposure parameters and the facet size. This follows from the modified heat transport in the bulk substrate during the ablation process, likely causing varying peak temperatures and surface tension. This made it difficult to develop an ablation pulse sequence for a desired mirror geometry using {}{a} reference glass, before applying it to an arbitrarily shaped pyramidal facet. {}{While some variability was observed in cavity mirror geometry across an array fabricated on reference glass, nominally attributed to laser power fluctuation and cross-talk between the translational axes of the flexure stage, the heat transport considerations suggested above were the dominant source of feature variation. Similar challenges were experienced in the fabrication of the closely spaced cavity arrays, where the creation of one mirror depression would modify the geometry of those ablated next to it.} Further, when facet sizes approached the 125$\mu$m diameter standard of single-mode fibre-tips, the anisotropy of the substrate around the ablation site was seen to induce minor feature warping. Since the warping had the tendency to modify mirror ellipticity, when formed into a cavity it has the potential to induce or alter birefringence~\cite{Uphoff:15}. This can limit the fidelity of remote entanglement generation in a quantum network~\cite{Kassa:20}. To create single-feature cavities with a mirror diameter similar to the substrate facet, restoring cylindrical symmetry to the facet may indeed be advantageous. However, as illustrated in Fig.~\ref{fig_facets}, expanding the facet size removes this issue. {}{This allowed the mirror features to be located such that their periphery was well separated from the substrate edge and neighbouring array mirrors, with nominally 100$\mu$m between ablation sites.}


\subsection{Measured Mirror Properties}
\label{sec:mirrors}
We experimentally verify the optical quality of the tapered mirrors and demonstrate reflective losses within standard limits for ablative processing and thus {}{suitability} for the construction of high-finesse cavities. Dielectric coating of the tapered substrates was performed by a commercial supplier (Laseroptik GmbH) using ion beam sputtering. The coating, consisting of quarter wave stacks of Ta$_2$O$_5$ and SiO$_2$, had a design wavelength of 780 nm and was specified for minimal transmission. To evaluate the mirror quality, we are interested in scattering losses arising from the features' surface roughness, absorption losses in the dielectric coating and residual transmission. Where possible, we ascribe these losses to aspects of mirror manufacturing process, acting as key feedback for future production cycles.

The mirror reflectivity was determined via the cavity finesse, using its common definition as the ratio of free spectral range to linewidth~\cite{Fox:13}:

\begin{equation}
    \mathcal{F} = {}{\frac{\pi(R_{i}R_{j})^{\frac{1}{4}}}{1-\sqrt{R_iR_j}} = \frac{2\pi}{\mathcal{L}_{i} + \mathcal{T}_{i}+\mathcal{L}_{j} + \mathcal{T}_{j}}} = \frac{\Delta\omega_{\textrm{fsr}}}{\Delta\omega_{\textrm{c}}},
\end{equation}

{}{where $R_{i,j}$ are mirror reflectivities, $\mathcal{L}_{i,j}$ are scattering and absorption losses summed for each mirror and $\mathcal{T}_{i,j}$ is their transmission}. $\Delta\omega_{\textrm{fsr}}$ is the free spectral range of the cavity and $\Delta\omega_{\textrm{c}}$ is the full-width at half maximum of its resonance line shape. This ratio is measured via laser spectroscopy, addressing the cavity with a laser while scanning its length. The laser wavelength was chosen to correspond to the design wavelength of the dielectric stack in order to minimise transmission losses. 
{The laser has a linewidth of a few hundred kilohertz, significantly smaller than the expected cavity linewidth.} Sidebands of frequency $\Delta\nu_{\textrm{SB}} = 100$ MHz were applied to the laser. The cavity length was modified by driving a piezoelectric transducer with a constant voltage gradient. Over short time intervals at the centre of the scan range it is reasonable to assume that the mirrors have a constant velocity. Therefore, $\Delta\omega_{\textrm{c}}$ can be determined from the ratio of the time taken to cross the main cavity resonance feature, $\Delta t_{2\kappa}$, to the time taken to move between resonance with the sidebands, $\Delta t_{\textrm{SB}}$:

\begin{equation}
    \Delta\omega_{\textrm{c}} = 2 \pi \cdot 2 \Delta\nu_{\textrm{SB}} \frac{\Delta t_{2\kappa}}{\Delta t_{\textrm{SB}}}.
\end{equation}

However, such an approach is not suitable for the determination of free spectral range owing to the nonlinearity of piezoelectric extension being significant over the translational distance of the free spectral range. In order to compensate for the non-linear expansion, we rely upon the resolution of higher order modes. For each mode, coordinates of $(t,2\Delta \nu_{\textrm{SB}}/\Delta t_{\textrm{SB}})$ are fitted, mapping the nonlinear rate of cavity resonance frequency change. The integral under this curve, taken between two consecutive fundamental mode resonances, then gives a robust measure of free spectral range. 


\begin{figure}[h!]
\centering\includegraphics[width=12cm]{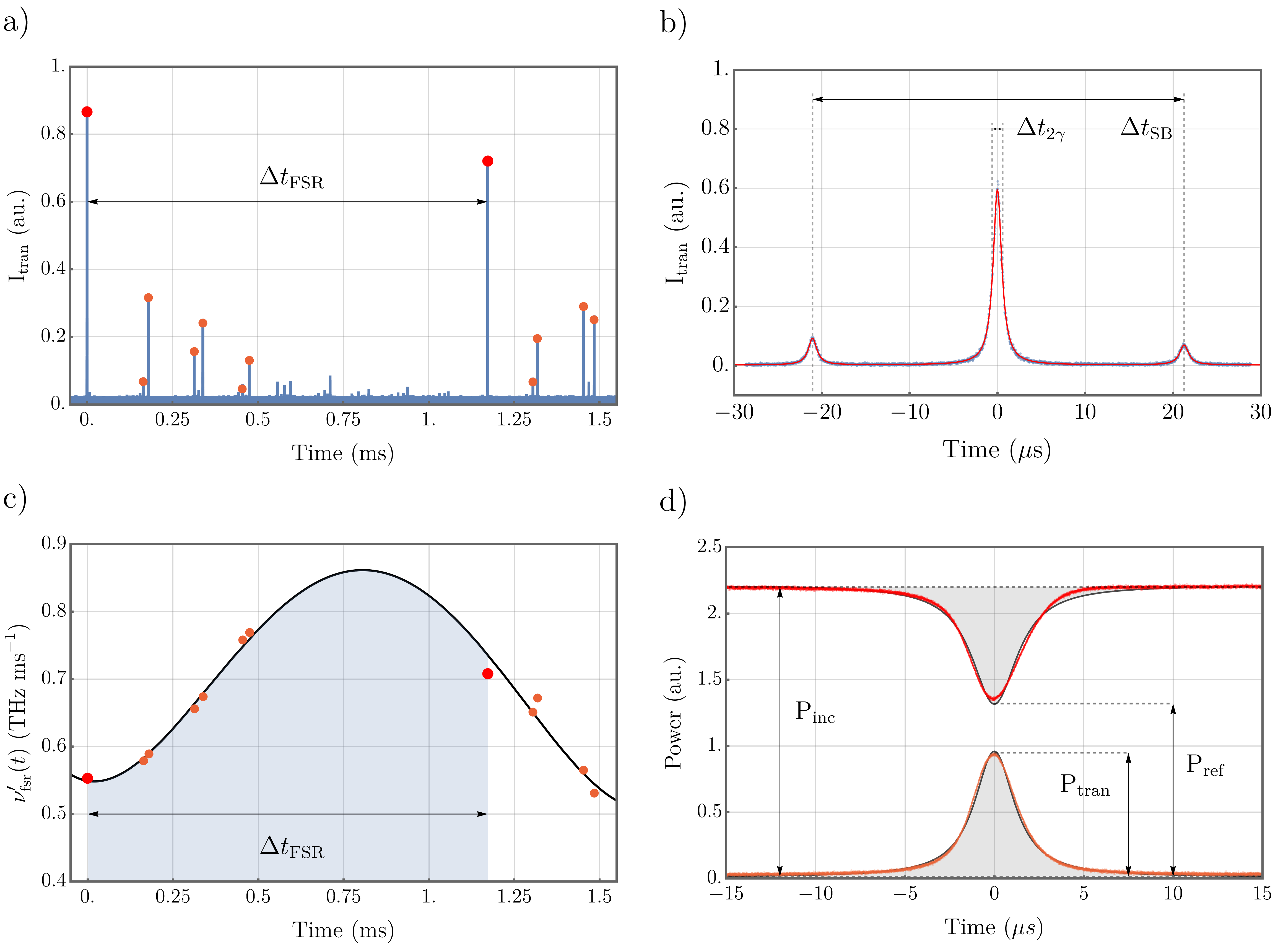}
\caption{Cavity characterisation process. a) $\Delta\nu_{\textrm{fsr}}$ was measured by addressing the cavity with a frequency stable laser and timing consecutive fundamental resonances as its length was scanned. b) A representative measurement of cavity linewidth, showing the transmission profile of a {}{$\Delta\nu_{\textrm{c}} = 5.64 (0.12) $ MHz} linewidth cavity illuminated by a probe beam with $\Delta\nu_{\textrm{SB}} = \pm100$MHz sidebands. In the time axis of changing mirror separation, these frequencies are represented by the FWHM resonance crossing time, $\Delta t_{2\gamma}$, and sideband crossing time, $\Delta t_{\textrm{SB}}$. c) To compensate for piezo nonlinearity, the frequency scanning rate was fitted for every mode, mapping mirror velocity. Spectral separations are the integral region, where {}{$\Delta\nu_{\textrm{fsr}} = 875 (14)$ GHz} and is highlighted. d) To derive the geometric mode matching and thus mirror transmission, the optical power transmitted by the cavity, $P_{\text{trans}}$, is compared to the input power, $P_{\text{inc}}$ and the cancellation of back reflected light, $P_{\text{ref}}$. {}{All errors are determined statistically.}}
\label{fig_reflect}
\end{figure}


{}{The properties of new mirrors are determined by measuring the finesse in this manner, pairing them with a reference mirror of known characteristics. An illustration of the geometry of this cavity is given as Figure~\ref{fig_pyramid}c). Two types of super-polished reference mirrors were used, coated for different transmission values, with each being of 50mm radius of curvature and optically characterised by the manufacturer (Research Electro-Optics, Inc.). Following the labelling convention of Figure~\ref{fig_facets}, pyramidal mirrors a) and b) were characterised using a reference mirror with $\mathcal{L}=2$ppm and $\mathcal{T}=0.5$ppm, giving finesse values of $146500\pm3400$ and $156500\pm7200$ respectively, corresponding to pyramidal losses of $\mathcal{L}_{a}+\mathcal{T}_{a}=40.4\pm1.0$ and $\mathcal{L}_{b}+\mathcal{T}_{b}=37.7\pm1.8$. Pyramidal mirrors b) and the top right feature of c) were characterised using a reference mirror with $\mathcal{L}=2$ppm and $\mathcal{T}=38.3$ppm, giving finesse values of $79300\pm4100$ and $77000\pm3800$ respectively, corresponding to pyramidal losses of $\mathcal{L}_{b}+\mathcal{T}_{b}=38.9\pm4.1$ and $\mathcal{L}_{c}+\mathcal{T}_{c}=41.3\pm4.0$. All errors are calculated statistically.} 

{}{To determine the transmission of the pyramidal mirror coating and thus isolate their incoherent loss rate, we compare {}{the} transmitted optical power from the cavity to {}{the} back-reflected power on resonance, following the technique of~\cite{Hood:01}. For pyramidal mirror b) paired with the high transmission reference mirror, this gave $\mathcal{T}_{b} = 1.8\pm 0.4$ppm. This highlights that low transmission rates are achievable for the presented design, which may be useful in experiments where directionality of the photonic emission is desired. This can be achieved by constructing the cavity with one mirror with low transmission and another mirror with high transmission, engineered by simply modifying the number of layers in the dielectric stack. Assuming each mirror possesses similar transmission, from the above characterisations we can therefore estimate that the pyramidal mirror possesses scattering and absorption losses of $35.9\pm1.8$ppm. These losses outweigh the transmission substantially, which is a well-known issue in most cavity implementations using laser-ablated or FIB machined mirrors. However, in combination with low-loss superpolished mirror of 40ppm transmission, our new cavities do nonetheless transmit about half the light from the cavity to outside free-running modes. }


\begin{figure}[h!]
\centering\includegraphics[width=12cm]{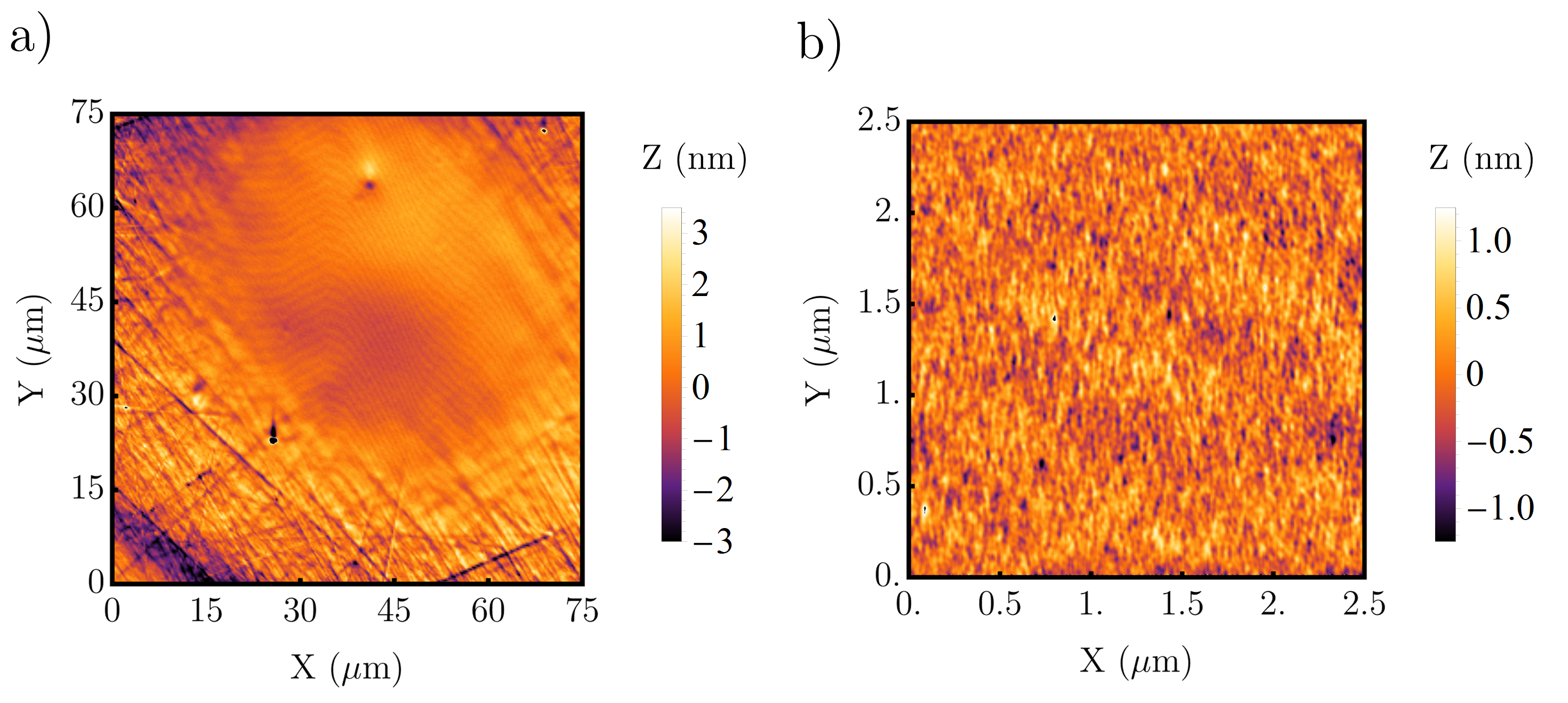}
\caption{Surface heating by laser ablation. a) The topography of an ablated pyramidal mirror is shown, measured using an AFM. Striations and surface damage from tip polishing can be seen in the periphery of the image. Towards the site of ablation, (45$\mu$m, 45$\mu$m), these surface deviations have been smoothed by thermal annealing. b) A 2.5$\mu$m$\times$ 2.5$\mu$m region at the centre of the ablation feature is shown, from which, surface roughness can be determined. This was measured to be 0.33(0.01) nm, with a polynomial fit applied to remove feature curvature.}
\label{fig_afm}
\end{figure}


 {}{In an ideal cavity array, the mirror surfaces have parallel normal axes, such that when paired with an equivalent substrate, the optical modes sit at the centre of each mirror. Any misalignment of these axes, caused by error in ablation beam pointing or underlying curvature of the initial substrate facet, may cause the cavity modes to migrate towards the mirror periphery. To ensure that any such error does not degrade cavity finesse, good optical quality should be achieved across the entire mirror. In combination with atomic force microscopy (AFM) analysis, surface homogeneity was indicated in the finesse measurements described above, via the constant linewidth of several of the higher order modes.}

Prior to the application of {}{a} reflective coating, scattering losses were {}{predicted} by AFM analysis (NaniteAFM, Nanosurf). This highlighted the important role of thermal annealing, clearly visible in Fig~\ref{fig_afm}a). RMS surface roughness was determined over a {}{central} area of 2.5$\mu$m $ \times$ $2.5\mu$m as $\sigma_{rms}=0.33\pm0.01$ nm (see Fig.\ref{fig_afm}b)). {}{Using a demonstrated model~\cite{Bennett:92}, this corresponds to scattering losses of $29\pm1$ ppm, leaving approximately 7ppm to be associated with coating absorption. Absorption of cavity locking light can be known to generate thermal instability of an optical cavity around resonance, especially for FFPCs, but no such effect was observed in the probing of cavity linewidth performed above. Overall, these values are competitive with other laser machined mirrors, for which losses in the range $\mathcal{L} = 20-35$ ppm were achieved~\cite{Garcia:18,Hunger:10,Saavedra:21,Gallego:18,Brekenfeld:20}.} {}{In realising an optical quality and geometrical range similar to FFPCs, we outline scope to achieve the favourable atom-cavity coupling parameters observed in such systems. }
{}{For the $S_{1/2} \Longleftrightarrow P_{3/2}$ transition of Rubidium atoms we estimate the cooperativity to range from 60 for cavities made from mirrors with radius of curvature of 1200 $\mu$m to 150 for cavities made from mirrors with radius of curvature of 500 $\mu$m (assuming a Finesse of 78,000). For the $P_{1/2} \Longleftrightarrow D_{3/2}$ transition of Ca$^+$ ions, the cooperativities are estimated to range from 5 and 11 for the respective cavities.}
  
 
\section{Cavity arrays}\label{sec:MultiFeatures}

To mediate a coherent interaction between an atom and a cavity, one must normally tune the cavity resonance to a transition line of the atom. From a technical perspective, this condition may be expressed as a requisite cavity length, {}{with a stability requirement} approximated by $\frac{\lambda}{2 \mathcal{F}}$. For the high-finesse cavities used in strong-coupling experiments, this limit is easily on the order of picometres. The common approach to achieving operational stability is to mount the cavity mirrors on piezoelectric transducers, locking the cavity length to a frequency stable laser via the Pound-Drever-Hall technique~\cite{Black:01}. To prevent the reference laser from inducing atomic scattering, it is often far detuned from atomic resonances.

\begin{figure}[h!]
\centering\includegraphics[width=12cm]{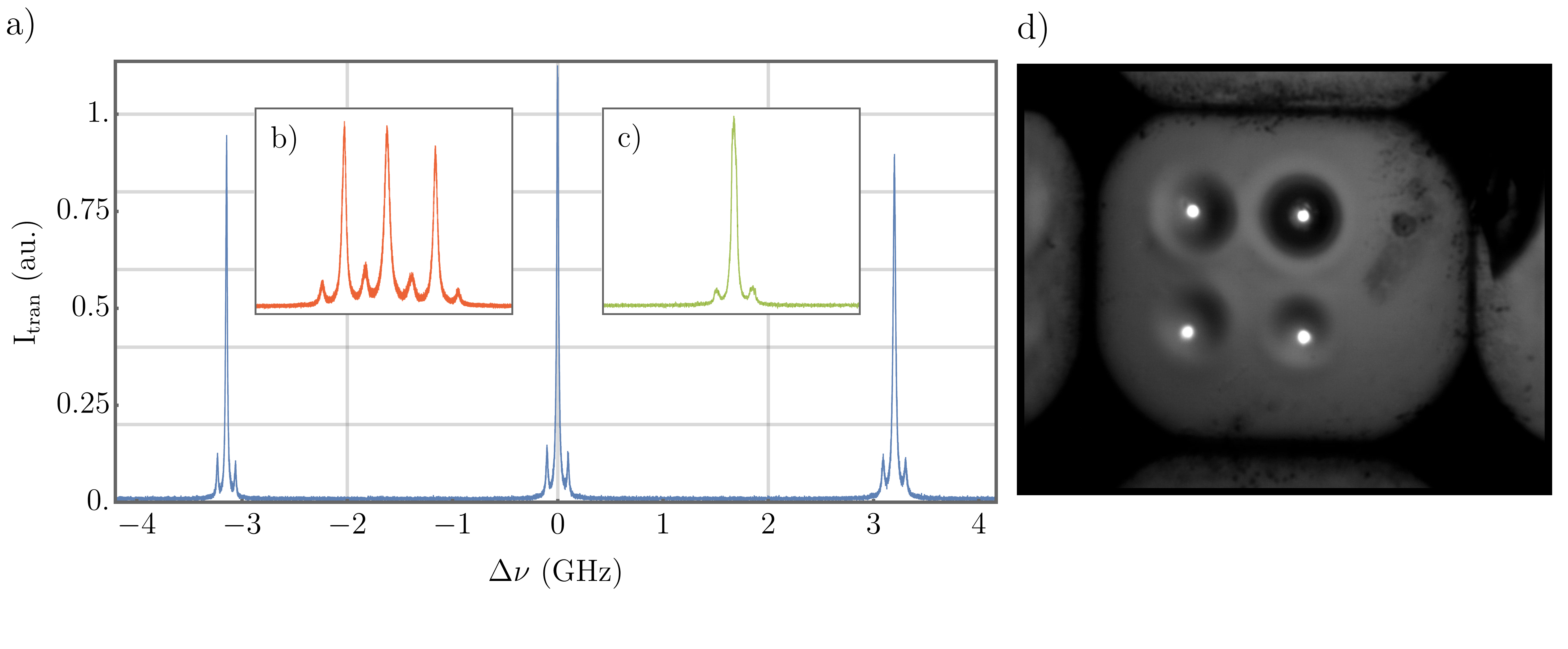}
\caption{Simultaneous operation of a 2$\times$2 cavity array. The joint transmission profile of three cavities during a cavity length scan is shown. The frequency separation of the resonances for the three cavities is controlled by adjusting the tip and tilt angles of the near-planar reference mirror such that the three cavity lengths can be tuned (see main text). To demonstrate control over their respective frequency detunings, traces are shown for separations of a) $|\Delta\nu| = 3.15$~GHz b) $|\Delta\nu| = 200$~MHz c) $|\Delta\nu| = 0$~MHz. d) An optical image of the cavity modes as the cavity length is scanned, demonstrating coupling of each cavity to the fundamental TEM mode. The cavity in the top right is actually several GHz detuned, but appears to be co-resonant with the others owing to the long integration time of the camera used to capture this image.}
\label{fig_array}
\end{figure}

{}{To realise co-resonance of multiple cavity modes to a single optical frequency, and thus simulate co-resonance to an atomic transition, an expansion was made to the} experimental architecture of Section 2.3. Here, four cavities are formed from a single bulk substrate hosting four mirror facets, and a near-planar reference substrate, addressed with four independently steerable {}{branches of the frequency stable laser}. The translation of the near-planar reference mirror in the cavity axis brings each cavity mode into visible resonance, initially at discordant {}{positions}. To control the relative frequency offset between each cavity resonance, the near-planar mirror is rotated in tilt and yaw to change the relative length of the cavities in the array. A rough optimisation of the near-planar mirror angle is performed by bringing each cavity mode towards the centre of the ablated features (by using a camera feed imaging the position of the cavity modes as in Fig.\ref{fig_array}d)). Fine tailoring of the angles is achieved using piezoelectric actuators {}{while monitoring} the resonance signal from a photodetector (Fig. \ref{fig_array}a)-c)). A set of demonstrations are given of the control of the frequency of three cavity mode resonances, with fixed spacing. {}{To integrate such a scheme with trapped atoms or ions would require the use of an in-vacuum piezoelectric stage, allowing for curvilinear translation in-situ. While this entails an engineering complexity over that of simple shear piezos for length tuning, the use of similar stages is already in practice for FFPCs~\cite{Teller:22}. In operation, this stage may also allow to compensate for slow drifts in cavity alignment.}

{}{As three points define a plane, the fourth cavity mode cannot be brought into perfect resonance using curvilinear translations. Increasing the number of co-resonant modes would require manufacturing mirror depths to an accuracy of several picometers; well outside the ability of current glass shaping methods. While only a modest number of co-resonant cavity modes increases the utility as an atom-photon interface, for increased scalability one can combine the co-resonant cavities with quantum shift registers, such as re-configurable atom or ion traps, to eventually use cavity-mediated entanglement schemes to produce larger cluster states as a resource for quantum information processing.}

{}{Finally, as an expansion to the scheme presented above, cavity length stabilisation may be performed by optical illumination of just one reference cavity from the array, with mode detuning compensated by appropriate frequency shift of the locking laser. For conventional cavity locking
schemes which apply continual feedback, the absence of laser light in the atom-resonant cavities would provide benefit in reducing off-resonant scattering rates and unwanted AC stark shifts. Long term stability of such a locking scheme could not be evaluated in the current apparatus, due to the characterisation being performed at atmospheric pressure and the presence of inherent drifts in the 6-axis flexure stage employed.}

\section{Conclusion}

We have developed and characterised an open-access optical cavity array on a single pair of mirror substrates, with mirror templates manufactured using laser ablation. 
{}{We measured losses of {$\mathcal{L} \approx 36$ppm} and combined the multi-facet mirror with a super-polished mirror having a transmission  rate nearly equal to the losses of the machined substrate, thus allowing access to about half of the photons in the cavity mode.}
Towards the creation of a multi-emitter cavity quantum interface, we have demonstrated simultaneous laser characterisation of four cavity modes, formed between a pair of substrates. By tuning their respective resonance frequencies with piezoelectric-actuated curvilinear translations, three cavities were placed in simultaneous co-resonance with the addressing laser.

This open-access cavity combines three key advantages of other implementations that were hitherto not simultaneously realised. The first is that it allows the preservation of the spatial mode of the cavity output, allowing for enhanced collection efficiencies as well as full polarisation control. The second is that the macroscopic nature of bulk substrates facilitates the suppression of vibration and effective heat dissipation. This significantly alleviates challenges in stabilising a high finesse cavity's length to an atomic transition frequency, being perturbed either by mechanical vibration or absorbed heat from the locking laser~\cite{Brachmann:16}. The third is the achievement of a low radius of mirror curvature, which gives rise to small cavity mode volumes and extremely strong coupling.

A limitation in using the tapered substrates, which have different facet dimensions, is that the laser ablation procedure needs to be optimised individually owing to a varying heat dissipation rate which is highly dependent on the facet geometry. The rectangular shape of the facets also required additional consideration to compensate for ellipticities that would come about due to the absence of cylindrical symmetry. These issues, however, could be resolved in future implementations by employing selective laser etching techniques \cite{Masajada:13,Ragg:19}, to produce substrates of regular dimensions, but also to carve ring cutouts with depths corresponding to the penetration depth of the CO$_2$ laser, in order to restore cylindrical symmetry at the ablation sites.

{}{The presented multi-cavity system could act as several nodes in a quantum network, scaling single photon sources~\cite{Kuhn:02} and quantum memories~\cite{Specht:11}. Or, it could be used to realise schemes which inherently require arrays of atom-cavity systems, such as those for distributed quantum computation~\cite{Duan:05,Alessio:06} or networking~\cite{Uphoff:16}. These multi-resonant cavity arrays open up pathways to explore new physics in cavity quantum electrodynamics, such as many-body quantum phenomena~\cite{Hartmann:08}.}

\section*{Acknowledgements}
The authors would like to acknowledge support for this work from the UK National Quantum Technologies Programme (NQIT hub, EP/M013243/1) and the EU-ITN LIMQUET. This work was supported by JST Moonshot R\&D [Grant Number JPMJMS2063].

\section*{Disclosures}

The authors declare that there are no conflicts of interest related to this article.

\section*{Data Availability}

{}{Data underlying the results presented in this paper are not publicly available at this time but may be obtained from the authors upon reasonable request.
}

\bibliography{PyrCav}

\end{document}